\pdfoutput=1

\documentclass[10pt,journal,compsoc]{IEEEtran}
\usepackage{amsmath}

\usepackage{bm}

\ifCLASSOPTIONcompsoc
  \usepackage[nocompress]{cite}
\else
  \usepackage{cite}
\fi
 \usepackage{graphicx}

\ifCLASSINFOpdf
\else
\fi
\begin{document}
%
\title{Better Than Reference In Low Light Image Enhancement: Conditional Re-Enhancement Networks}
%
%
%
%

\author{Yu~Zhang,
        Xiaoguang~Di,
		Bin~Zhang,
        Ruihang~Ji,
	 and~Chunhui~Wang 
\IEEEcompsocitemizethanks{\IEEEcompsocthanksitem  This work was supported by the National Natural Science Foundation of China (No. 61775048) and the Fundamental Research Funds for the Central Universities, China (No. ZDXMPY20180103)
\IEEEcompsocthanksitem Y. Zhang, B. Zhang and C. Wang are with the Department of Electronic Science and technology, Harbin Institute of Technology, Harbin 150001, China.(E-mail: hitzhangyu@qq.com; teamup@yeah.net; wang2352@hit.edu.cn)
\IEEEcompsocthanksitem X. Di and R. Ji are with the Department of Control Science and Engineering, Harbin Institute of Technology, Harbin 150001, China.(E-mail: dixiaoguang@hit.edu.cn; jiruihang@hit.edu.cn)}
\thanks{}}

%
%

\markboth{}%
{Yu \MakeLowercase{\textit{et al.}}: Better Than Reference In Low Light Image Enhancement: Conditional Re-Enhancement Network}
%



\IEEEtitleabstractindextext{%
\begin{abstract}
\renewcommand{\raggedright}{\leftskip=0pt \rightskip=0pt plus 0cm}
\raggedright
Low light images suffer from severe noise, low brightness, low contrast, etc. In previous researches, many image enhancement methods have been proposed, but few methods can deal with these problems simultaneously. In this paper, to solve these problems simultaneously, we propose a low light image enhancement method that can combined with supervised learning and previous HSV (Hue, Saturation, Value) or Retinex model based image enhancement methods. First, we analyse the relationship between the HSV color space and the Retinex theory, and show that the V channel (V channel in HSV color space, equals the maximum channel in RGB color space) of the enhanced image can well represent the contrast and brightness enhancement process. Then, a data-driven conditional re-enhancement network (denoted as CRENet) is proposed. The network takes low light images as input and the enhanced V channel as condition, then it can re-enhance the contrast and brightness of the low light image and at the same time reduce noise and color distortion. It should be noted that during the training process, any paired images with different exposure time can be used for training, and there is no need to carefully select the supervised images which will save a lot. In addition, it takes less than 20 ms to process a color image with the resolution 400*600 on a 2080Ti GPU. Finally, some comparative experiments are implemented to prove the effectiveness of the method. The results show that the method proposed in this paper can significantly improve the quality of the enhanced image, and by combining with other image contrast enhancement methods, the final enhancement result can even be better than the reference image in contrast and brightness. (Code will be available at https://github.com/hitzhangyu/image-enhancement-with-denoise)
\end{abstract}

}

\maketitle

\IEEEdisplaynontitleabstractindextext

%
\IEEEpeerreviewmaketitle

\begin{figure*}[htbp]
\centering
\includegraphics [width=\linewidth]{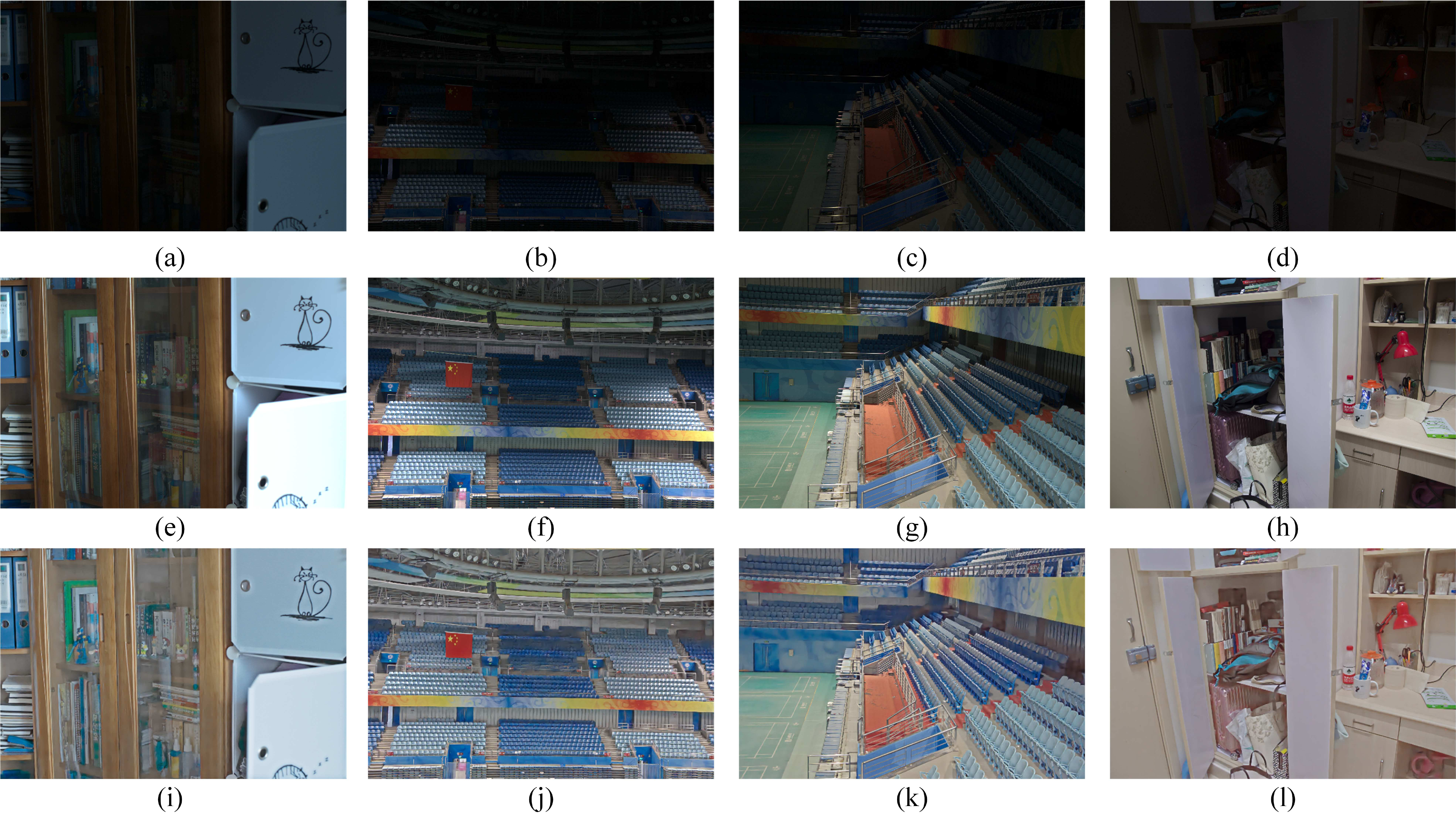}
\caption{Visual Comparison with Reference. \textbf{Top Row}: low-light images. \textbf{Middle Row:} Reference images. \textbf{Bottom Row:} The results enhanced by our method combined with Gamma correction.
}
\label{fig1}
\end{figure*}

\IEEEraisesectionheading{\section{Introduction}\label{sec:introduction}}

%
%
%
%
\IEEEPARstart{W}{hen} the environment light is low, such as at night or in a dark room, captured images will be low light images. This kind of images always suffer from low contrast, low brightness, serious noise, and so on. Low light image enhancement method is used to solve these problems before high-level tasks. In the past decades, researchers have proposed lots of non-learning based image enhancement methods, such as \cite{jobson1997multiscale},\cite{kimmel2003variational}, \cite{guo2016lime}, etc. Recently, with the development of deep learning, many supervised and unsupervised learning based image enhancement methods have been proposed, such as \cite{wei2018deep}, \cite{zhang2019kindling}, \cite{zhang2020self}, \cite{xiong2020unsupervised}, etc., and achieve promising results. However, whether these methods are based on learning or not, they have not been able to solve all these problems in low light images well simultaneously.

\par For non-learning based methods \cite{guo2016lime},\cite{pizer1987adaptive},\cite{lee2013contrast}, most of them can significantly improve the image contrast and brightness. However, it is difficult for these methods to reduce or suppress noise directly, and they may even amplify noise or cause color distortion during the enhancement process. The later denoising operation often brings problems such as blur,  disappearance of details, etc. For unsupervised image enhancement works, since it is difficult to introduce some prior knowledge to the learning pipeline, especially noise and color terms, there is always a problem of noise and color distortion \cite{xiong2020unsupervised}. For most supervised works \cite{zhang2019kindling},\cite{chen2018learning}, there must be a hyperparameter which can be used to connect the input and reference image during training, and this is caused by non-one-to-one correspondence between input and reference images. The hyperparameter can be used to adjust the contrast and brightness of the whole image during test, however, it is hard to obtain the hyperparameter automatically, and the brightness and contrast of some local areas may not be satisfactory even if the hyperparameter is adjusted carefully. 

\par In this paper, we provide a low-light image enhancement framework that can integrate unsupervised learning or non-learning methods with supervised learning methods to solve the low contrast, low brightness, noise and color distortion in low light images simultaneously.

\par The effectiveness of supervised methods often depends on the quality of the data. However, in low-level image processing tasks, it is difficult to get input/label pairs like high-level tasks, especially in image enhancement tasks. Although we can get some images with different illumination from the same scene by changing the lighting or modifying the camera parameters, we cannot guarantee that the reference image has good contrast and brightness on each image block (e.g. Fig. 1 (e)-(f)). At the same time, as one low light image can correspond to many high light images, it is hard to ensure consistency in the selected data (similar input image blocks should correspond to similar reference image blocks in light). In fact, it can be considered that there is no ground-truth image here. Then the problem can be summarized as how we can train the networks without the unique or the best ground-truth image and how we can connect the input image and reference image when they do not meet the consistency condition during training. Different from the way of introducing a single hyperparameter such as time ratio \cite{chen2018learning} into the enhancement process in previous supervised works, we propose a new framework which can use point-wise parameters related to contrast to achieve training and the parameters can be automatically obtained through other contrast enhancement methods during test. It can be seen in Fig. 1, combining our method with some contrast enhancement methods, like gamma correction, we can adjust the contrast of the enhanced image and reduce the noise simultaneously.  And contrast in some local areas of the enhanced images are even better than those in the reference images, such as bookcase, seats in the dark part of stadium. 

\par As with some previous works \cite{Lore2015LLNet}, we divided the low light enhancement problem into two sub-problems, one is the contrast and brightness enhancement problem, and the other is the denoising and color restoration problem. Different from the previous method to solve the two sub-problems separately, we use contrast enhancement methods to generate point-wise contrast and brightness proposals and use them as a condition to re-enhance the low light image, and at the same time reduce the noise and color distortion. The method in this paper can be combined with any image contrast and brightness enhancement method based on HSV color space or Retinex model. And the network can be trained without the need of carefully selecting reference images, and any paired images with different light can be used for training.

\par Our contributions can be summarized as follows:
\begin{itemize}

\item A conditional re-enhancement network(denoted as CRENet) for low light image enhancement is proposed, which can solve the low contrast, low brightness, noise and color distortion simultaneously. The CRENet can be combined with existing image enhancement method and use the enhanced V channel as a condition to achieve re-enhancement of low-light images. In this process, the CRENet can maintain the contrast and brightness of other enhancement method, while reducing noise and color distortion at the same time.
\item Compared with other learning-based methods, the hyperparameters included in our method are point-wise and are directly related to image contrast and brightness, which makes it possible that the enhanced image has better contrast than the reference image through adjusting the brightness and contrast of the local area of the image, and that is hard for other learning-based methods with only one hyperparameters.

\item By combining with other contrast and brightness enhancement methods, the proposed method does not need to carefully select the reference images with good exposure, and any paired images with different brightness can be used for training.

\end{itemize}

\begin{figure}[htbp]
\centering
\includegraphics [width=3.5in]{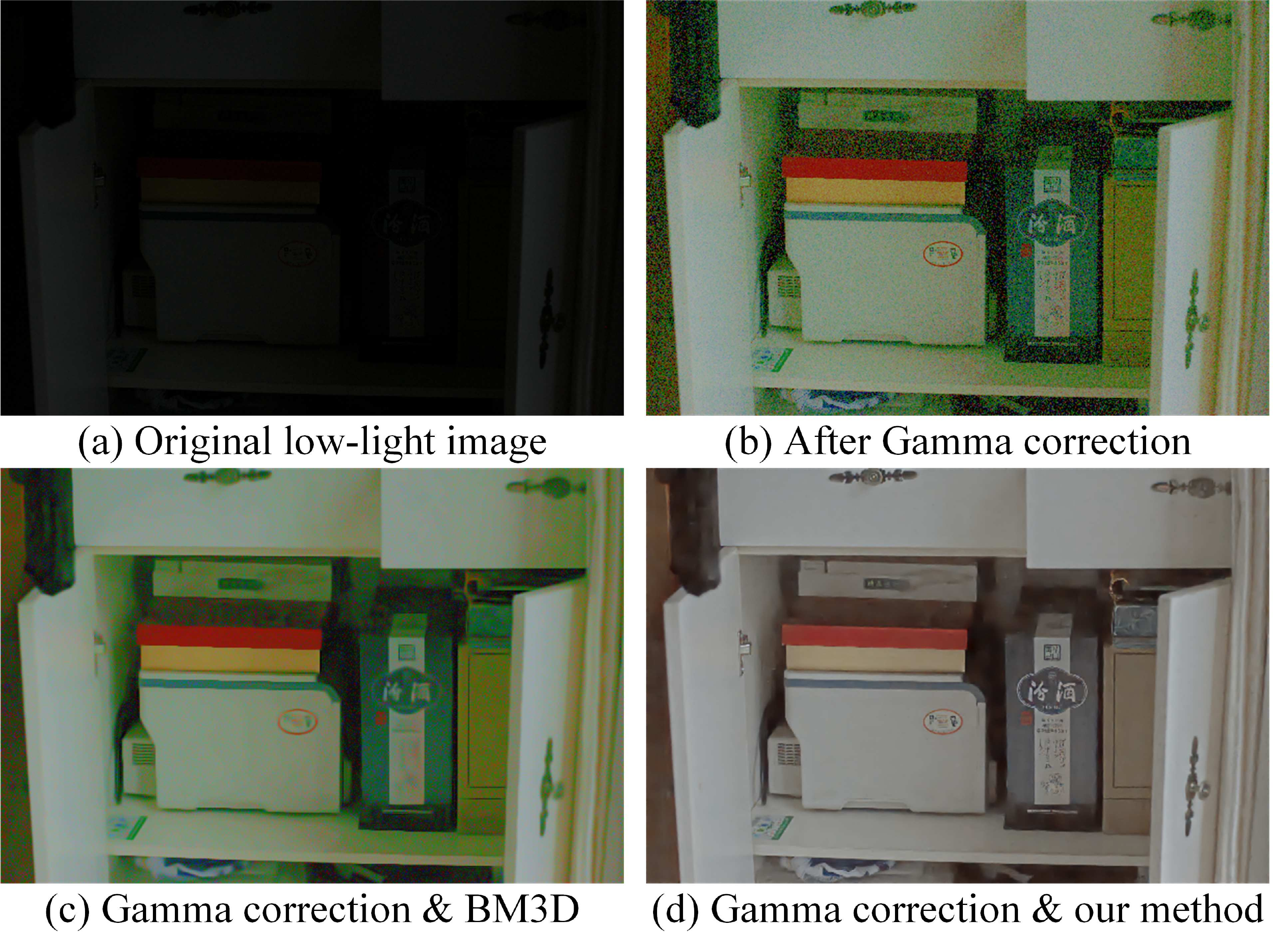}
\caption{Visual Comparison with BM3D.  (Best viewed on high-resolution displays with zoom-in.)
}
\label{fig2}
\end{figure}

\section{Related Works}
\textbf{A.Non-learning based image enhancement methods}

\par For non-learning based single low light image enhancement methods, there are mainly histogram equalization, methods based on dehazing or Retinex model and other improved methods based on those methods. 

\par Histogram Equalization(HE) is one of the most wildly used methods, however, it cannot avoid the problems of detail disappearance, poor color restoration, noise amplification and so on. Although many improved methods have been proposed to solve those problems \cite{pizer1987adaptive},\cite{lee2013contrast}, \cite{pisano1998contrast}, \cite{kim1997contrast}, \cite{wang1999image}, \cite{celik2011contextual}, there are still many problems in applying histogram equalization directly to image enhancement.

\par Dong et al.\cite{dong2011fast} proposed the method based on dehazing model firstly and then some studies extended these works \cite{li2015low}, although these methods have achieved some good effect, they lack corresponding physical model, which limits the application of the method in various scenes. 

\par Some works based on Retinex model are proposed to maintain image details and naturalness \cite{guo2016lime}, \cite{wang2013naturalness}, \cite{fu2016fusion}, however, the denoising process before or after the enhancement process will still cause blur or loss of details. To solve the noise and hole effect, and preserve more details, many algorithms based on the variational Retinex model are proposed and achieved good results, such as \cite{kimmel2003variational},  \cite{fu2013variational}, \cite{park2017low}, \cite{fu2019hybrid}, however, most of them will cost too much time due to the need of multiple iterations to solve the variational equation.

\par Most of the non learning methods focus on contrast and brightness enhancement, and then use general denoising methods (like BM3D \cite{dabov2007image}) and white balance to remove noise and correct the color, however, those methods always bring blur and cannot solve the problem of color distortion. As shown in Fig. 2, although Gamma correction improves the contrast and brightness of the image, after the denoising operation of BM3D, some details disappeared. 

\noindent \textbf{B.Learning based image enhancement methods}
\par As we know, LLNET [37] is the first work to use deep neural networks to solve image enhancement problem, and it proposes to train the networks with synthetic noisy and dark images separately, but it does not consider the characteristics of the natural images. Some other methods also use synthetic data sets, such as \cite{li2018lightennet}, \cite{yang2016enhancement}, \cite{shen2017msr}, although the data obtained by these methods seems to look like low light images, it is difficult to truly reflect some characteristics of low light images, such as noise, color distortion, overexposed and underexposed areas existed in the same image, etc.

\par Chen et al. \cite{chen2018learning} introduce a dataset which contains real raw low light images and corresponding raw high light images for training, and they introduce an amplification ratio to connect the input and reference images and multiply the ratio to a certain layer in the network. The ratio is set to be the exposure time difference between the input and reference images during training. This method can solve the problem of noise and color distortion well, however, the ratio must be chosen by user during test which limit the widely use of this method, and it can not solve the inappropriate contrast problem well. It can be seen that this algorithm provides an exposure time adjustment method instead of physically adjusting the exposure time, it is obvious that only by adjusting the exposure time cannot work well in the night scene, there will still be some over-enhanced (saturated) areas or under-enhanced areas in the image. In our previous work \cite{fu2020learning}, we provide a way to automatically learn the expected exposure time, but still can not solved the problem of contrast. Because we can never provide the ground-truth reference images with proper contrast in every local area, and without other constraints, it is difficult to solve this problem. 

\par In \cite{wei2018deep}, it is proposed to introduce the Retinex model into the training process to connect the reflection images of input and reference. However, due to the lack of constraints on the noise of the reflection image, additional denoising methods(like BM3D \cite{dabov2007image}) still need to be introduced. The method proposed in \cite{zhang2019kindling} looks like the combination of \cite{chen2018learning} and \cite{wei2018deep}. Compared to \cite{wei2018deep}, \cite{zhang2019kindling} provides an extra brightness ratio to connect the illuminantion images of input and reference and adds a subnet called restoration-net to achieve denoising. However, during the test, it need to manually adjust the ratio parameter to obtain better enhancement results. Although these methods use real low light data for training, they do not constrain the contrast of the enhanced image, which caused the problem of over-enhancement (saturation) or under-enhancement in the enhanced image.

\par In our previous work \cite{zhang2020self}, we provide a max entropy Retinex model to achieve self-supervised learning and at the same time constrain the contrast during training. However, due to the lack of strong constrains on color and noise, the enhanced image still looks like a night one in color and the noise cannot be removed well, and during test, we cannot guarantee the well contrast on every local area.

\par Xiong et al. \cite{xiong2020unsupervised} decompose the low light image enhancement task into two stages: contrast enhancement and noise removal, and propose an unsupervised framework. However, in the absence of constraints on the image contrast, the contrast and brightness after image enhancement may be still unsatisfactory, and in this paper we also prove that the Retinex model used in their first stage cannot ensure the color information close to the real day image.
\par Although these deep learning based methods have achieved good visual effects in low light image enhancement, they still cannot solve the low contrast, low brightness, noise and color distortion simultaneously. 


\section{Method}
\subsection{Relationship between HSV color space and Retinex model}
\par Recently, a lot of low-light image enhancement works are based on the following Retinex model:

\begin{equation}
\label{eqn1}
\mathbf{F}=\mathbf{R}\circ \mathbf{I}
\end{equation}

\noindent where $\mathbf{F}$ and $\mathbf{I}$ represent the captured image and the illumination image respectively, $\mathbf{R}$ represents the reflection or the desired image, $\circ$ represents the element-wise multiplication. Most of the works assume that the three color channels of the image have the same illumination in order to simplify the model \cite{wei2018deep},\cite{zhang2019kindling}, and the maximum value of the three color channels is generally used as the initial estimate of the illumination map \cite{guo2016lime}. In the following description, we refer to this simplified Retinex model as the Retinex model. Through a simple transformation, it can be proven that image enhancement methods based on this simplified Retinex model are equivalent to make  enhancement operation on V channel in HSV color space and remain the H and S channels unchanged.
\par The color image can be divided into three channels according to the value of each pixel in RGB color space:
\begin{equation}
\label{eqn2}
\left\{\begin{matrix}\mathbf{L}\left ( x \right )=\max\limits_{c\in \left \{ R,G,B \right \}}\mathbf{F}^{c}\left ( x \right )
\\
\\\mathbf{M}\left ( x \right )=\mathop{median}\limits_{c\in \left \{ R,G,B \right \}}\mathbf{F}^{c}\left ( x \right )
\\
\\\mathbf{N}\left ( x \right )=\min\limits_{c\in \left \{ R,G,B \right \}}\mathbf{F}^{c}\left ( x \right )

\end{matrix}\right.
\end{equation}

\noindent where $x$ represents an individual pixel. Before the process of image enhancement, the captured image $\mathbf{F}$ in HSV color space can be expressed as follows:

\begin{equation}
\label{eqn3}
\mathbf{V}_{b}\left ( x \right )=\mathbf{L}\left ( x \right )
\end{equation}

\begin{equation}
\label{eqn4}
\mathbf{S}_{b}\left ( x \right )=\frac{\mathbf{L}\left ( x \right )-\mathbf{N}\left ( x \right )}{\mathbf{L}\left ( x \right )}
\end{equation}

\begin{equation}
\label{eqn5}
\mathbf{H}_{b}\left ( x \right )=c_{1} +c_{2} \frac{\mathbf{M}\left ( x \right )-\mathbf{N}\left ( x \right )}{\mathbf{L}\left ( x \right )-\mathbf{N}\left ( x \right )}
\end{equation}

\noindent where $c_{1}$ can be 60, 120, 240, $c_{2}$ can be $\pm$60, their values depend on the three color channels of the image, and for one image, $c_{1}$ and $c_{2}$ are certain value in each pixel. $\mathbf{V}_{b}$, $\mathbf{S}_{b}$, $\mathbf{H}_{b}$ represent the V, S and H channel of the low light image before enhancement, respectively, and the subscript $b$ represents before. 

\par Based on the Retinex model, the desired image $ \mathbf{R}$ can be obtained by the following formula:

\begin{equation}
\label{eqn6}
 \mathbf{R}\left ( x \right )=\mathbf{F}\left ( x \right )/\left ( \mathbf{I}\left ( x \right )+\varepsilon  \right )
\end{equation}
\noindent where $\varepsilon$ is a very small constant to avoid the zero denominator. For simplicity of writing, we omit $\varepsilon$ in the following formula and directly use $\mathbf{I}\left ( x \right )$ to represent $ \mathbf{I}\left ( x \right )+\varepsilon $.

\par According to Equation (6), the enhanced image $\mathbf{R}$ in HSV color space can be expressed as follows:

\begin{equation}
\label{eqn7}
 \mathbf{V}_{a}\left ( x \right )=\mathbf{L}\left ( x \right )/\mathbf{I}\left ( x \right )= \mathbf{V}\left ( x \right )_{before}/\mathbf{I}\left ( x \right )
\end{equation}

\begin{equation}
\label{eqn8}
\mathbf{S}_{a}\left ( x \right )=\frac{\mathbf{L}\left ( x \right )/ \mathbf{I}\left ( x \right )-\mathbf{N}\left ( x \right )/I\left ( x \right )}{\mathbf{L}\left ( x \right )/ \mathbf{I}\left ( x \right )}=\mathbf{S}_{b}\left ( x \right )
\end{equation}

\begin{equation}
\label{eqn9}
\mathbf{H}_{a}\left ( x \right )=c_{1} + c_{2} \frac{\mathbf{M}\left ( x \right )/ \mathbf{I}\left ( x \right )-\mathbf{N}\left ( x \right )/ \mathbf{I}\left ( x \right )}{\mathbf{L}\left ( x \right )/ \mathbf{I}\left ( x \right )-\mathbf{N}\left ( x \right )/ \mathbf{I}\left ( x \right )}= \mathbf{H}_{b}\left ( x \right )
\end{equation}
\noindent where $\mathbf{V}_{a}$, $\mathbf{S}_{a}$, $\mathbf{H}_{a}$ represent the V, S and H channel of the low light image after enhancement, respectively, and the subscript $a$ represents after.

\par It can be seen that H and S channels of the captured image and the enhanced image are the same, and the enhancement operation only works on the V channel, then the enhanced V channel can well represent the image enhancement operation for methods based on HSV or Retinex model. It is obvious that the hue and saturation are different between the image at night and day, also different between the low light and high light images. This is caused by the non-linearity of camera response curve, and even different color channel have different response curve, so methods based on the simplified Retinex model cannot ensure that the color of the enhanced images looks like a real image captured at day or high light. 

\par Therefore, we provide a supervised image enhancement method called the Conditional Re-Enhancement Network (CRNet) which takes the enhanced V channel as an additional point-wise parameter to allow the network focus on learning the H and S channels which change with V channel. At the same time, with the enhanced V channel as additional parameter, we can benefit from previous researches \cite{guo2016lime}, \cite{zhang2020self},  \cite{pizer1987adaptive} on image contrast and brightness enhancement, and the supervised learning also shows well performance in noise suppression. In this way, we can simultaneously solve the problems of low contrast, low brightness, noise, and color distortion in low light images.
\par Next, we provide the details of the network and training, and further explain the reason and rationality of using the V channel instead of a single parameter.

\begin{figure}[htbp]
\centering
\includegraphics [width=3.5in]{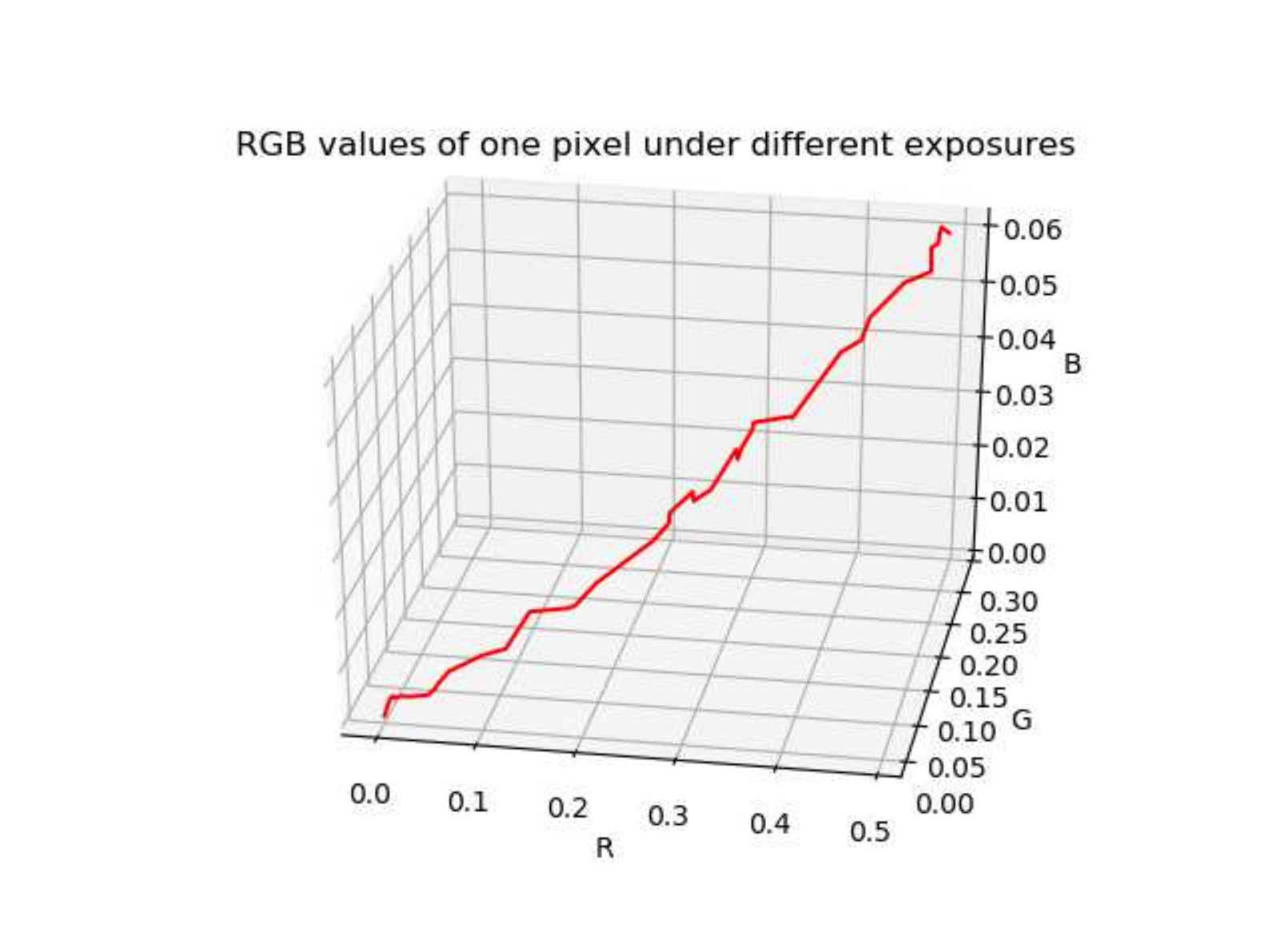}
\caption{RGB values of one pixel under different exposures.
}
\label{fig3}
\end{figure}

\subsection{Conditional re-enhancement network}
\par According to the camera response curve, a pixel on the input and reference image can be expressed as follows:
\begin{equation}
\label{eqn10}
\left\{\begin{matrix}
\mathbf{X}_{cij}=G_{c}\left ( \mathbf{E}_{i}\Delta t_{j} \right )
\\
\mathbf{Y}_{cik}=G_{c}\left ( \mathbf{E}_{i}\Delta t_{k} \right ) 
\end{matrix}\right.
\end{equation}
\noindent where $\mathbf{X}$ and $\mathbf{Y}$ represent the input low light image and the reference image respectively, $c$ represents one of the color channels, $G_{c}$ represents the camera response curve of $c$ channel, $i$ represents the pixel, $\Delta t_{j}$ and  $\Delta t_{k}$ represent the exposure time of low light image and high light image respectively, $\mathbf{E}_{i}$ represents the irradiance. Obviously, different reference images may have different exposure times for better visual effects, therefore most supervised works should include at least one time-related item $\alpha$ in order to connect the input and reference image.

\par Then, most of the learning based models can be described as follows:
\begin{equation}
\label{eqn11}
\max  p\left ( \theta |f_{\theta},\mathbf{Y},\textbf{X},\alpha \right )
\end{equation}

\noindent where $f$ and $\theta$ represents the enhancement network and parameters in the network respectively, and $p$ represents the probability. $\alpha$ represents the hyperparameter related to the time difference, which is used to connect $\mathbf{X}$ with manually selected $\mathbf{Y}$, and previous works adopted time ratio $\alpha=\Delta t_{k} / \Delta t_{j}$ \cite{chen2018learning}, average brightness ratio $\alpha=mean \left(\mathbf{Y} / \mathbf{X} \right)$ \cite{zhang2019kindling}, etc. As we mentioned before, there are two obvious problems in previous works. Firstly, it is difficult to assign a value to $\alpha$ automatically during application phase whether it is time or average brightness difference, because those values are not directly related to image contrast. Secondly, a single parameter can only enhance the whole brightness of the image, and can not guarantee a well contrast of the image on every local area. It can be illustrated through the following Equations (12) and (13).
\begin{table}
	\centering
	\caption{Conditional re-enhancement network structure}
	\resizebox{0.48\textwidth}{!}{
		\begin{tabular}{cccccc}
			\hline
			Inputs & Operator& Kernel  &Output Channels & Stride & Output Name \\ 
			\hline
			RGB\&Enhanced V 			& Conv\&ReLU   & $3\times3$ & 32  & 1 				& Conv0 \\
			RGB\&Enhanced V     		& Conv         & $9\times9$ & 64  & 1 				& Conv \\
			Conv            			& Conv\&ReLU   & $3\times3$ & 64  & 1 				& Conv1 \\
			Conv1           			& Conv\&ReLU   & $3\times3$ & 128 & 2$\downarrow$   & Conv2 \\
			Conv2           			& Conv\&ReLU   & $3\times3$ & 128 & 1 				& Conv3 \\
			Conv3           			& Conv\&ReLU   & $3\times3$ & 64  & 2$\uparrow$     & Conv4 \\
			Conv4\&Conv1    			& Concat       & -          & 128 & - 				& Conv5 \\
			Conv5		  					& Conv\&ReLU   & $3\times3$ & 64  & 1 				& Conv6 \\
			Conv6\&Conv0    			& Concat       & -          & 96  & - 				& Conv7 \\
			Conv7           			& Conv         & $3\times3$ & 64  & 1 				& Conv8 \\
			Conv8 		   					& Conv   & $3\times3$ & 3   & 1 					& Conv9 \\
			Conv9           			& Sigmoid      & -          & 3   & -				& Enhanced \\
			\hline
		\end{tabular}
	}
	\label{tab:structure}
\end{table}
\par According to Bayes rule, by calculating the negative logarithm of Equation (11), the training phase can be expressed as follows:
\begin{equation}
\label{eqn12}
\max p\left ( \theta |f_{\theta},\mathbf{Y},\mathbf{X},\alpha \left ( \mathbf{X},\mathbf{Y} \right ) \right )=\min_{\theta}||
f_{\theta}\left (\mathbf{X}, \alpha \left ( \mathbf{X},\mathbf{Y} \right ) \right )-\mathbf{Y}
||
\end{equation}

\par If we assume that $\mathbf{Y}$ is the optimal reference in a series of images with different exposure time and we can manually select $\alpha$, the best result during testing without other prior losses can be expressed as follows:

\begin{equation}
\label{eqn13}
\hat{\textbf{Y}}=f_{\theta}\left(\mathbf{X},\alpha\left ( \mathbf{X},\mathbf{Y} \right )\right )=\mathbf{Y}
\end{equation}

\noindent where $\hat{\mathbf{Y}}$ represents the enhanced image. It can be seen that the best result of the network is hard to exceed the manually selected reference $\mathbf{Y}$. 
In addition, it is obvious that in many low light scenes, we can never get the optimal image as a ground-truth reference by adjusting the exposure time (e.g. Fig. 1 (e)-(f)), which means that we cannot obtain a satisfactory image by only adjusting a single parameter $\alpha$.

\par In fact, without other prior, the end-to-end supervised method cannot solve the problem of low contrast in low light images, or even the problem of low brightness, unless we manually adjust the reference images carefully or use multi images to synthesize reference images. But this will also bring other problems, such as increased time cost, the adjusted image can not guarantee the consistency of brightness (the same low light image patches correspond to different reference images), etc. In order to solve the above problems, we propose the CRENet to explicitly control the contrast and brightness of the enhanced image. The CRENet takes the V channel enhanced by other image contrast enhancement methods as a point-wise condition, and can re-enhance the contrast and brightness of low light image according to the condition. 

\par The V channel is the max of the three color channels at each pixel, so the V channel of the reference image can be expressed with camera response curve as follows:

\begin{equation}
\label{eqn14}
\mathbf{V}_{ik}=G_{V}\left(\mathbf{E}_{i} \Delta t_{k} \right) 
\end{equation}

\par We make the same reasonable assumption that the function $G$ is monotonically increasing \cite{debevec2008recovering}. Therefore, for a fixed pixel $i$, there is  a certain $\Delta t_{k}$ with given $\mathbf{V}_{ik}$, which means that V channel is enough to connect other two channels between the input and reference during training since the three color channels have the same exposure time. As shown in Fig. 3, with the exposure changes, the R, G, B value at one pixel\footnotemark[1] can form a curve in 3D space which is related to the camera response curve. Our motivation is to let the network learn the curve, and then with a given V value, it can obtain the values of the other two channels (HSV and RGB are just different color representations, so we directly implement it in the RGB space when designing and training the network). Then, the V channel can be treated as a point-wise parameter that we can achieve different levels of enhancement to different areas, and through the Equation (2) to (7), it can be seen that the V channel can well represent the processing results of other contrast enhancement methods.

\footnotetext[1]{The data comes from real images taken under 40 different exposure conditions. Under each exposure condition, we collected 80 images and averaged them to reduce the noise.}
\begin{figure}[htbp]
\centering
\includegraphics [width=\linewidth]{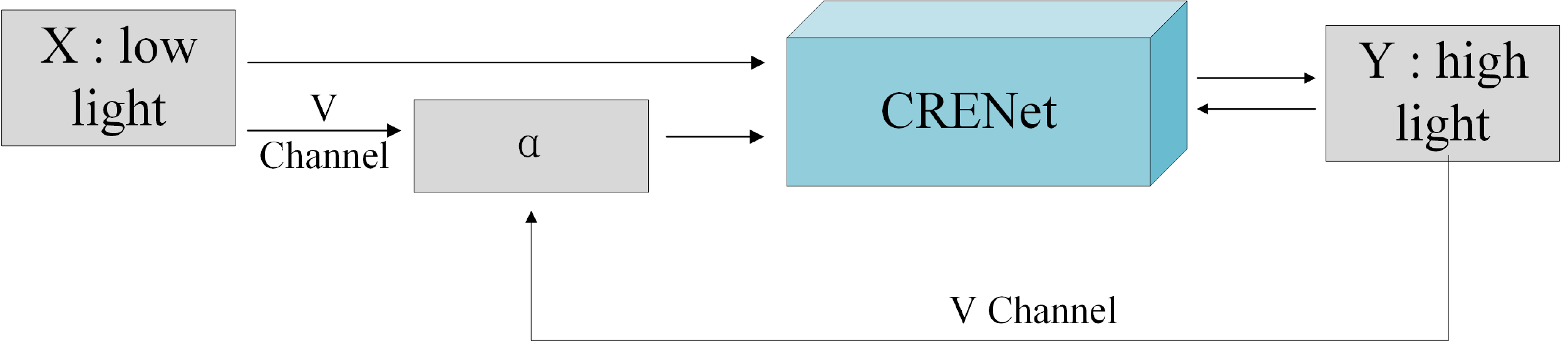}
\caption{ The structure of our training pipelines.
}
\label{fig4}
\end{figure}

\par During training, we can take the V channel of the reference images as point-wise parameter $\alpha$, then it can be expressed as follows:

\begin{equation}
\label{eqn15}
\mathbf{\alpha}=\max\limits_{c\in \left \{ R,G,B \right \}} \mathbf{Y}^{c} + \mathbf{n}
\end{equation}

\noindent where $\mathbf{n}$ represents the Gaussian noise, it is introduced in order to avoid the identity transformation and at the same time simulate the noise in V channel during testing.

\par Meanwhile, we can perform brightness mapping on the V channel of low light images according to the reference to get $\alpha$, then $\mathbf{\alpha}$ can be expressed as follows:

\begin{equation}
\label{eqn16}
\mathbf{\alpha}_{i}=w \max\limits_{c\in \left \{ R,G,B \right \}} \mathbf {X}_{i}^{c}
\end{equation}

\noindent where $w$ is the average brightness ratio which is calculated on every local area $\Omega$ around pixel $i$, and it can be expressed as follows:

\begin{equation}
\label{eqn17}
 w=\frac{\sum\limits_{x\in \Omega\left ( i \right ) }\mathbf{Y}\left ( x \right )}{\sum\limits_{x\in \Omega\left ( i \right ) }\mathbf{X}\left ( x \right )}  
\end{equation}

\par During testing ,$\mathbf{\alpha}$ can be expressed as follows:

\begin{equation}
\label{eqn18}
\mathbf{\alpha}=\max\limits_{c\in \left \{ R,G,B \right \}} g\left( \mathbf{X}^{c}\right) 
\end{equation}

\noindent where $g$ represents any contrast and brightness enhancement methods in HSV color space or based on Retinex model, such as histogram equalization, LIME \cite{guo2016lime} and self-supervised methods \cite{zhang2020self}, etc.

\par The training procedure is achieved by minimizing the loss between the enhanced image and the supervised image, and the loss can be expressed as follows:
\begin{equation}
\label{eqn19}
L=||\hat{\mathbf{Y}}-\mathbf{Y}||_{1}+{\rm SSIM} \left(\hat{\mathbf{Y}},\mathbf{Y} \right)
\end{equation}



\noindent where SSIM represents the structural similarity measurement \cite{wang2004image}. We have also test some other loss functions, such as perceptual loss, color loss expressed by outer product like \cite{xiong2020unsupervised}, loss in H and S channels in HSV space, gradient loss \cite{zhang2019kindling},etc., but the effect of these loss functions on the results is not obvious on LOL dataset, neither in visual effects nor in quantitative metrics.





\begin{table*}[htbp]
	\centering
	\caption{Quantitative comparison on LOL dataset in terms of NIQE, PSNR, and SSIM. Ori shows the results of original method}
	\begin{tabular}{c|cccccccc}
		\hline
		Metrics 		&	 &LAHE  &GAMMA  &LIME\cite{guo2016lime} &RetinexNet\cite{wei2018deep} &KinD\cite{zhang2019kindling} &Self-supervised\cite{zhang2020self} &Zero-DCE\cite{guo2020zero} \\
		\hline
			     		&ori   	&13.51	&10.36	&9.13		&9.73		&3.89		&3.72		&8.22	 \\
		NIQE     		&BM3D 	&8.50		&4.08		&4.03		 &4.61	&4.06		&4.15		&4.10	\\  
					&ours  	&\textbf{3.75}	&\textbf{3.68}	&\textbf{3.63}		&\textbf{3.68}		&\textbf{3.60}		&\textbf{3.63}		&\textbf{3.50}	 \\
		\hline
					&ori   	&12.88	&21.95	&21.06	 &21.63	&\textbf{25.82}	&24.37	&21.68\\
		PSNR   		&BM3D  	&15.35	&21.99	&22.22	 &22.41	&25.69	&24.64	&22.65\\
		    			&ours  	&\textbf{24.92}	&\textbf{24.83}	&\textbf{24.48}	&\textbf{24.81}	&25.43	&\textbf{24.76}	&\textbf{25.11}\\	
		\hline
		        		&ori   	&0.32		&0.62		&0.60		 &0.61	&\textbf{0.84}		&0.75 	&0.62	 \\
		SSIM        	&BM3D  	&0.41		&0.61		&0.64 	&0.64		&0.83		&0.76		&0.64	\\
					&ours 	&\textbf{0.80}		&\textbf{0.80}		&\textbf{0.80}  	&\textbf{0.80}		&0.83		&\textbf{0.80}		&\textbf{0.82}	\\
		\hline

	\end{tabular}
	\vspace{10pt}

	\label{table2}
\end{table*}

\begin{figure*}[htbp]
\centering 
\includegraphics [width=\linewidth]{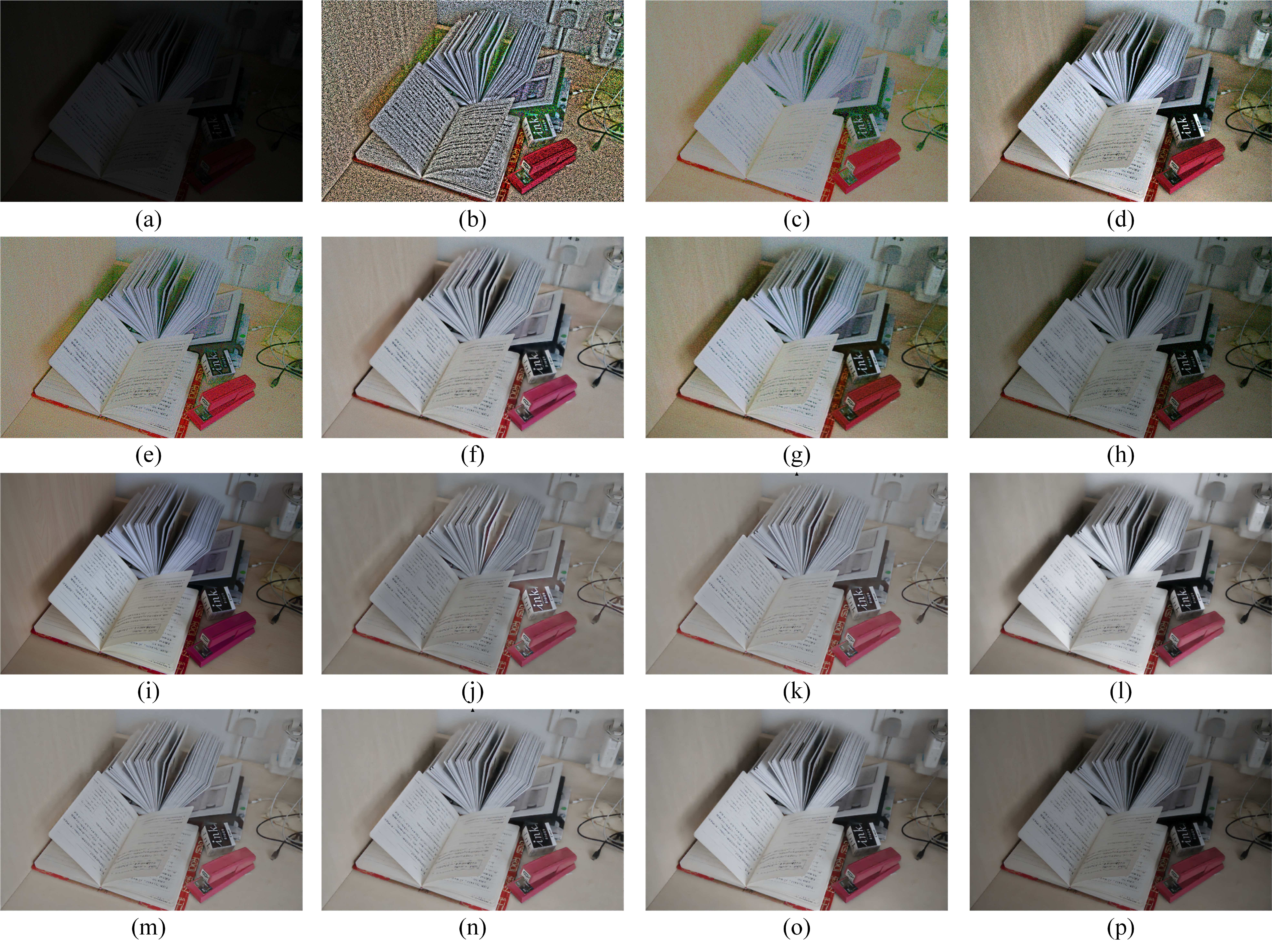}
\caption{The results enhanced by some methods and re-enhanced by our method. (a) Original. (b)-(h) are LAHE, Gamma correction, LIME \cite{guo2016lime}, RetinexNet \cite{wei2018deep}, KinD \cite{zhang2019kindling}, Self-supervised \cite{zhang2020self}, Zero-DCE \cite{guo2020zero}, respectively. (i) Reference. (j)-(p) are the re-enhanced results by our method corresponding to (b)-(h).
}
\label{fig5}
\end{figure*}

 \par The architecture of our CRENet is shown in Table 1 and the training pipeline is shown in Fig. 4. It should be noted that our method does not have special requirements for the network architecture. The network can be simplified or more carefully designed to reduce running time or further improve processing effect. We have tried complicated network, and it is found that a more complex network structure may bring better visual effects. However, this is not the focus of this paper, so we do not show relevant results in the experiment part.

\section{Experiment}
\subsection{Implementation Details}
%
%

\par The LOL (Low Light) dataset \cite{wei2018deep} which contains 500 image pairs is used for training and testing, 485 image pairs of the database are used for training and images size is 400x600. During the training process, the batch size is set to 48 and the patch size is set to 48x48. We use Adam stochastic optimization \cite{kingma2014adam} to train the network and the update rate is set to 0.001. The training and testing of the network are implemented on a Nvidia GTX 2080Ti GPU and Inter Core i9-9900K CPU, and the code is based on the tensorflow framework.

\subsection{Performance Evaluation}
\par\par In this part, we show the results of combining with some previous methods, including some methods under the HSV color space, such as LAHE (Local Area Histogram Equalization) and Gamma correction, and some methods based on Retinex model, such as LIME\cite{guo2016lime}, self-supervised image enhancement\cite{zhang2020self}, Retinex-Net\cite{wei2018deep}, KinD\cite{zhang2019kindling}, etc.  Also we compared our method with BM3D in the effect of denoising. Three metrics are adopted for quantitative comparison, which are PSNR, SSIM, and NIQE \cite{mittal2012making}. NIQE is a non-reference image quality assessment method, which can evaluate the naturalness of the image and a lower value indicates better quality. While, PSNR and SSIM are reference image quality assessment methods, which indicate the noise level and the structure similarity between the result and the reference, respectively. Please note that SSIM and PSNR are mainly used to evaluate structure and noise, but brightness differences will have a serious impact on the evaluation of these metrics. In order to better evaluate the results generated by different methods, we first perform a local brightness mapping on the enhanced image to exclude the influence of brightness differences before using SSIM and PSNR, which is the same as in Equation (16). And the mapping operation acts on the V channel and keep the H and S channels unchanged.

\par Table 2 and Fig. \ref{fig5} show the experiment results on LOL dataset when combined with different image enhancement methods. In Table 2, it can be seen that for most of the existing methods, our method can significantly reduce the noise and improve the structure similarity, and make the result more natural.  And as shown in Fig. 5,  taking images enhanced by other methods as condition, the CRENet can keep similar contrast and brightness to those enhanced images, and at the same time the CRENet can reduce the noise and color distortion in those enhanced images.  Also, it should be noted that the CRENet does not bring obvious blur in the process of denoising, and even for the LAHE method which produces severe noise, the re-enhancement network can work well (e.g. Fig. 5 (b) and (j)).  For KinD\cite{zhang2019kindling} method, our method does not achieve the effect of improving the PSNR and SSIM, it is because that there has less noise and color distortion on LOL dataset after KinD\cite{zhang2019kindling}. However, KinD can be treated as a whole image exposure time adjustment method, it means that it can not ensure a proper contrast or brightness on every local area. As shown in Fig. 6, we have tried different ratio, such as 5 and 8, and the author suggested that the maximum is 5. It can be seen that the lower right corner area of the image cannot be enhanced well even we change the ratio. As a matter of fact, we even tried the case when the ratio parameter is 100 in the experiment, but still cannot get good results.  



%
%

\par Also, we study the influence of different source of $\alpha$ on training, the source of which are from the reference image with noise, the low light image after mapping, and the mixture of the two. The results are shown in Table 3 and Fig. 7. It can be seen that, when the condition $\alpha$ is from the mixture of the low light and high light images, the network show better results on the evaluation metrics.  In Fig. 7,  it can be found that when the $\alpha$  comes from the reference,  there are still some low frequency noise and impulse noise, such as Fig. 7 (b).  However, if the $\alpha$ comes from the low light images with brightness mapping operation, there are less noise.  This is because we only add Gaussian noise to the reference images in the process of obtaining $\alpha$, which cannot simulate the real noise in test process very well.  In fact, it also shows that the real low light image contains far more complex noise than Gaussian noise, which means that it is hard to obtain promising results in real low light image enhancement tasks with only synthetic low light images. Training with the real data will also bring some problems in previous works, such as the need of adjusting parameters during test and carefully selecting reference during training, and enhanced results are difficult to exceed the reference, however, the proposed method can well solve those problems.

\begin{table}[htbp]
	\centering
	\caption{The influence of different $\alpha$ during training on the evaluation metrics}
	\begin{tabular}{c|ccc}
		\hline
		$\alpha$ 					&NIQE	 &PSNR  &SSIM\\
		\hline
		$Reference + Nosie$				&4.37		&23.55	   &0.774\\
		$Low light + Mappping$			&3.83		&24.83	   &0.821\\
		$Mixture$						&\textbf{3.72}		&\textbf{24.96} 	   &\textbf{0.823}\\
		\hline

	\end{tabular}
	\vspace{10pt}

	\label{tab:condition}
\end{table}

\begin{figure}[htbp]
\centering
\includegraphics [width=3.5in]{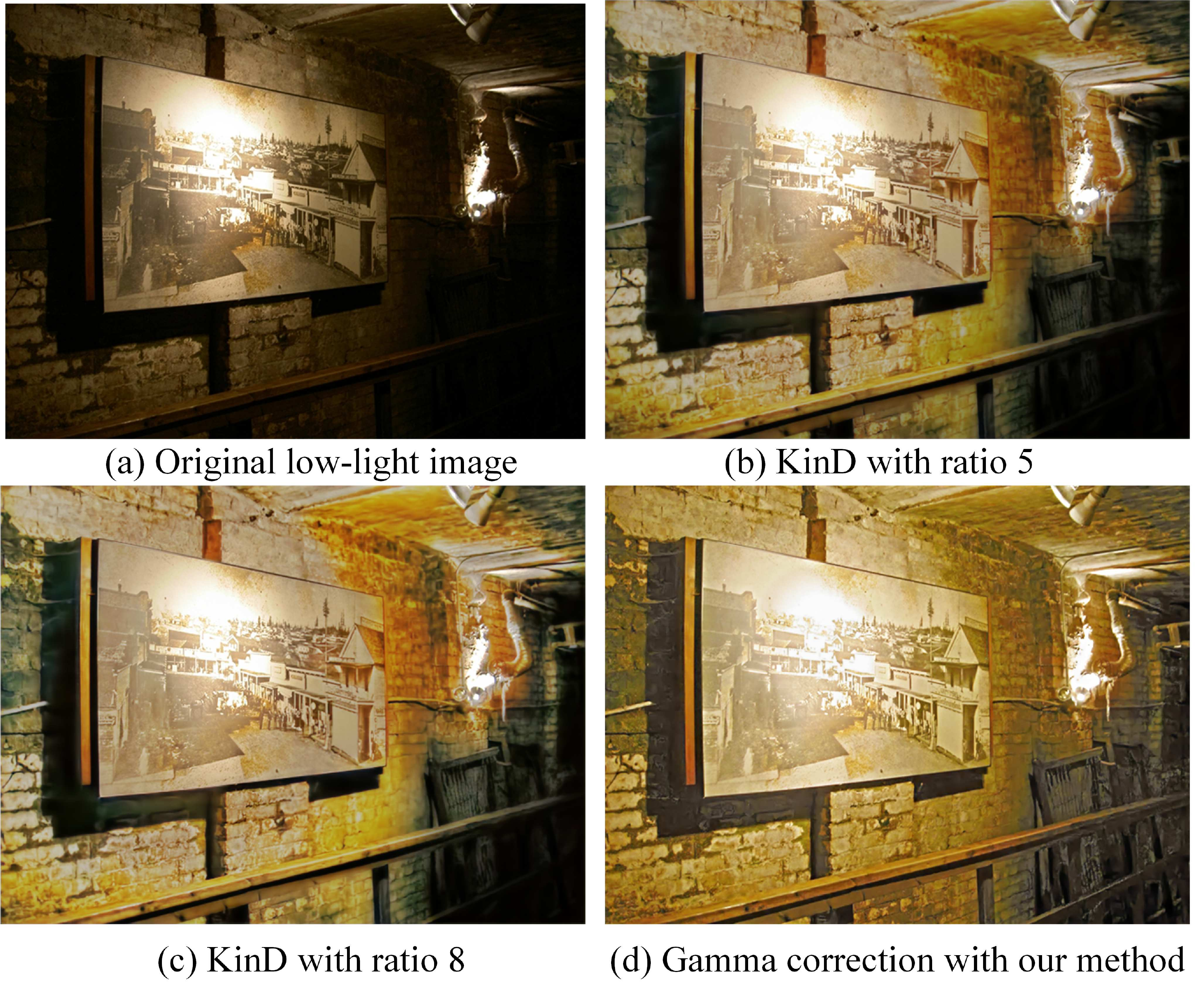}
\caption{Visual Comparison with single hyperparameter based method. (a) Original. (b) KinD\cite{zhang2019kindling} with ratio 5.(c) KinD \cite{zhang2019kindling} with ratio 8. (d) Gamma correction with our method.  (Best viewed on high-resolution displays with zoom-in.)  
}
\label{fig6}
\end{figure}

\par We also tested the consistency of the output results when the input images are the same scene with different brightness, as shown in Fig. 8. It can be seen that the enhanced results basically maintain the consistency of the brightness of the same blocks.

\begin{figure}[htbp]
\centering
\includegraphics [width=\linewidth]{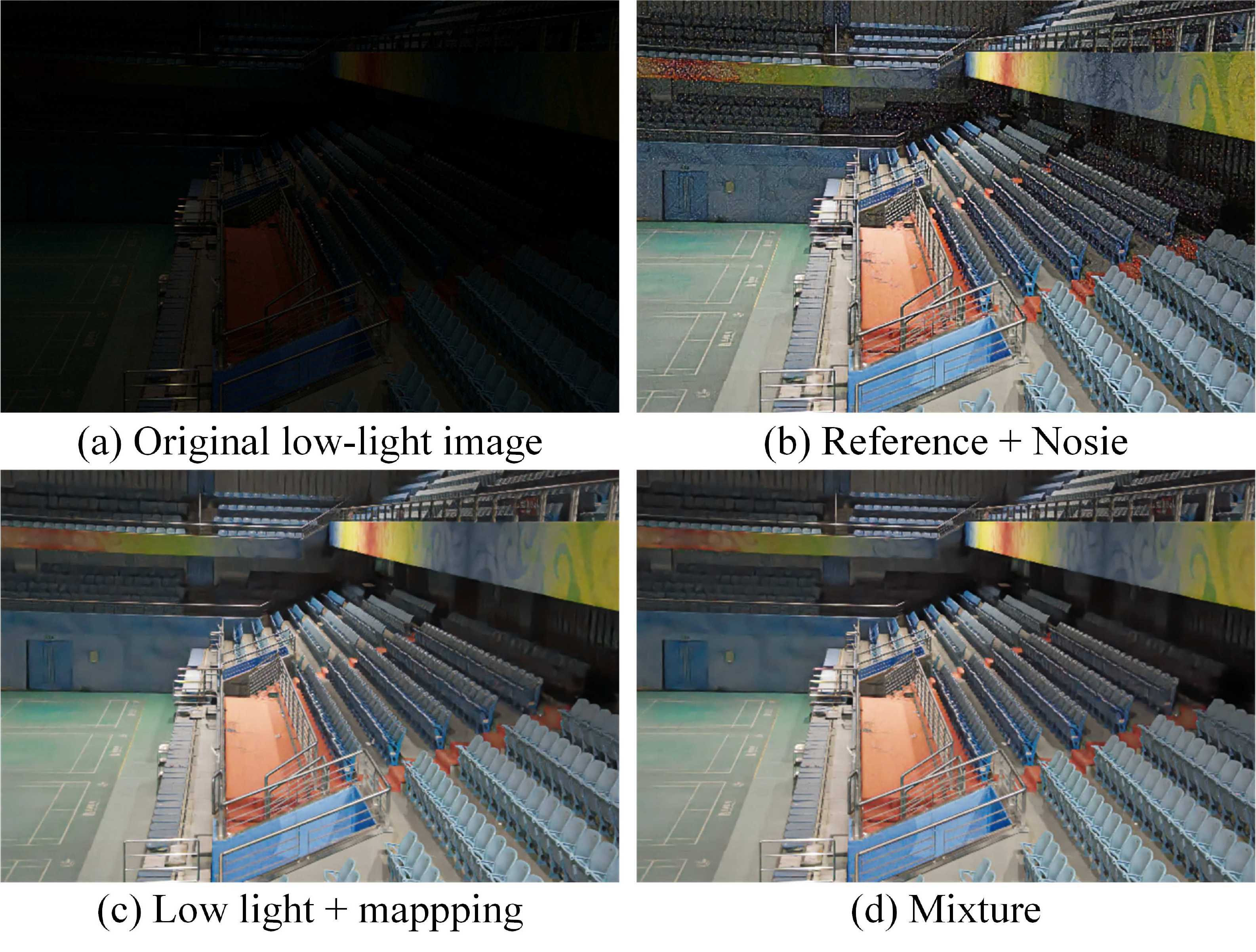}
\caption{Visual comparison of different  source of $\alpha$ during training.  (Best viewed on high-resolution displays with zoom-in.)
}
\label{fig7}
\end{figure}

\begin{figure}[htbp]
\centering
\includegraphics [width=\linewidth]{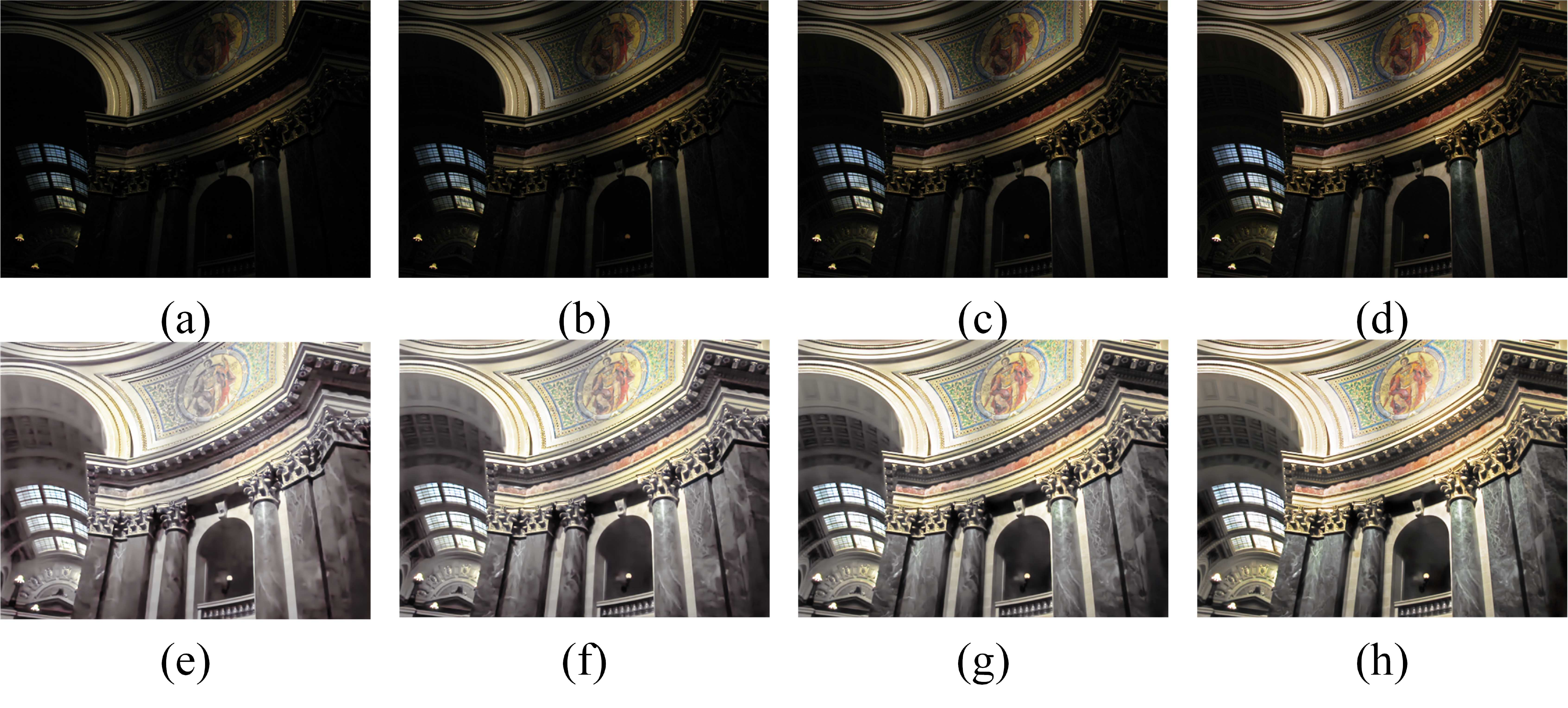}
\caption{Brightness consistency test results. \textbf{Top Row:} low light images with different exposure time. \textbf{Bottom Row:} the enhanced results by our method combined with HE.  (Best viewed on high-resolution displays with zoom-in.)
}
\label{fig8}
\end{figure}

\section{Conclusion}

\par In this paper, we propose a conditional re-enhancement network for low-light image enhancement to solve low contrast, low light, noise and color distortion simultaneously. The network can be combined with any contrast enhancement method which is based on HSV color space or simplified Retinex model, and enhanced V channel by other methods is treated as conditions to achieve re-enhancement. And the experiment results show the effectiveness and advantages of the method. Also, there is still some shortcomings: the final enhancement result depends on the contrast enhancement method, and the image saturation is reduced in some areas (e.g. Fig. 1 (l)). Future research will focus on applying the technology in long-term and night-time localization to eliminate the interference of light on feature extraction and generate corresponding night-time data sets, and using extremely low light images or raw images for training, etc.


%





\ifCLASSOPTIONcaptionsoff
  \newpage
\fi



%


\bibliographystyle{IEEEtran}
\bibliography{IEEEabrv,mybibfile}
%

\begin{IEEEbiography}[{\includegraphics[width=1in,height=1.25in,clip,keepaspectratio]{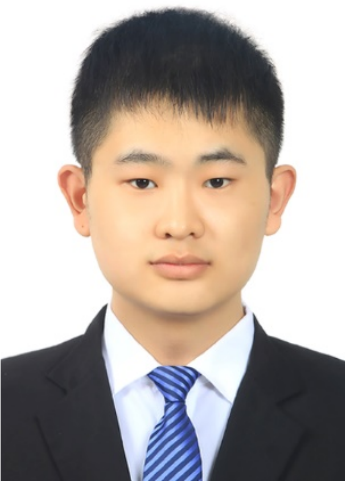}}]{Yu Zhang}
was born in Hebei, China. He received the B.E. degree and M.S.degree in Control Science and Engineering from Harbin Institute of Technology, China, in 2016 and 2018, respectively. He is currently studying for a Ph.D degree in Electronic science and technology in National Key laboratory of Tunable Laser Technology, Harbin Institute of Technology. His research interests include image restoration and enhancement, SLAM.
\end{IEEEbiography}

\begin{IEEEbiography}[{\includegraphics[width=1in,height=1.25in,clip,keepaspectratio]{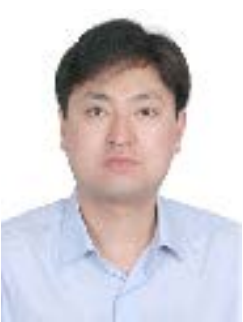}}]{Xiaoguang Di}
was born in Heilongjiang,China. He received the M.S. and Ph.D degree in navigation, guidance and control from Northwestern Polytechnical University, China, in 1999 and 2004, respectively. He is currently an associate professor with the Control and Simulation Center, Harbin Institute of Technology, Where he is in charge of courses in digital image processing and computer vision. His current research interests include real-time image restoration and enhancement, 3D object detection and recognition, SLAM. Prof. Di is a member of China Simulation Federation and Chinese Society of Astronautics.
\end{IEEEbiography}

\begin{IEEEbiography}[{\includegraphics[width=1in,height=1.25in,clip,keepaspectratio]{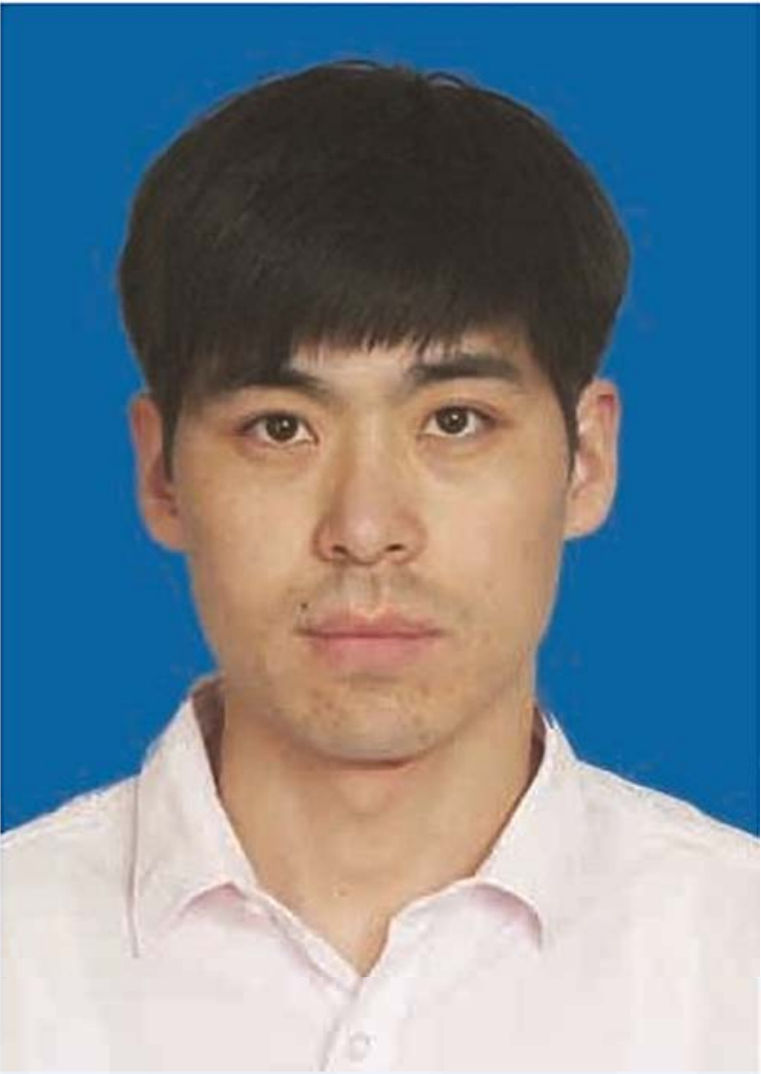}}]{Bin Zhang}
 was born in Gansu, China. He received the B.Sc. degree in apply physics and the M.Sc. degree in physics from China University of Petroleum, in 2010 and 2013, respectively. He is currently studying for a PhD degree in National Key Laboratory of Tunable Laser Technology, Harbin Institute of Technology. His research interests include laser imaging and laser transmission.
\end{IEEEbiography}

\begin{IEEEbiography}[{\includegraphics[width=1in,height=1.25in,clip,keepaspectratio]{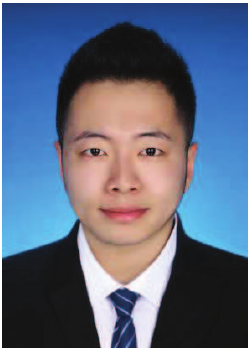}}]{Ruihang Ji}
(S’16) received the B.S. degree in
automation engineering from the Harbin Institute of Technology, Harbin, China, in 2016, where he is currently pursuing the Ph.D. degree in control science and engineering.

He is currently an exchange Ph.D. student supported by the China Scholarship Council with the Department of Electrical and Computer Engineering, National University of Singapore, Singapore. His current research interests include adaptive control, robust control, Unmanned Aerial Vehicle and computer vision.
\end{IEEEbiography}

\begin{IEEEbiography}[{\includegraphics[width=1in,height=1.25in,clip,keepaspectratio]{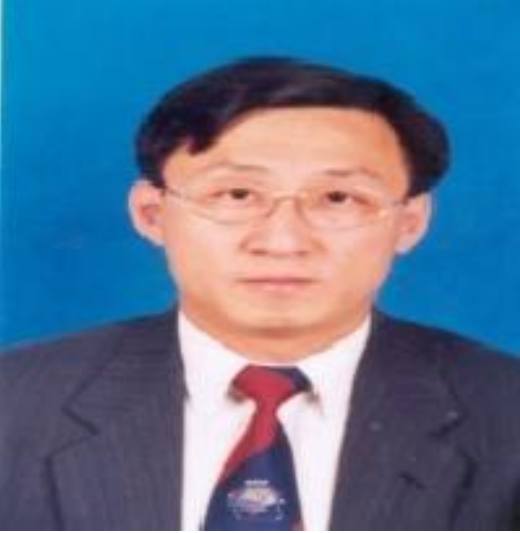}}]{Chunhui Wang}
Chunhui Wang received the BS, MS, and PhD degrees from Harbin Institute of Technology in 1987, 1991, and 2005, respectively. He is currently a processor and Deputy Director, Institute of Optoelectronic Technology, Harbin Institute of Technology. His current major research interests are laser remote sensing, lidar, laser detection and recognition.
\end{IEEEbiography}




\end{document}